\documentclass[fleqn,twoside]{article}[14pt]
\usepackage{epsf,multicol,ifthen}
\usepackage{ujp}
\usepackage[cp1251]{inputenc}
\usepackage[english,ukrainian]{babel}
\usepackage{amstext}
\usepackage{amssymb}
\usepackage{cite}
\usepackage {graphicx}
\usepackage{amsmath,latexsym,epsfig,epsf,rotate}

\date{}
\def\bt{{\mbox{$\beta$}}}
\def\sbt{{\mbox{$\beta^2$}}}
\def\spo{{\mbox{$p_0^2$}}}
\def\sro{{\mbox{$r_0^2$}}}
\def\sx{{\mbox{$x^2$}}}
\def\sl{{\mbox{$\lambda^2$}}}

\nazva{A SIMULTANEOUS CENTER--OF--MASS CORRECTION OF NUCLEON DENSITY AND MOMENTUM DISTRIBUTIONS IN NUCLEI \\
~~\\ 
\small{\texttt{(accepted to Ukr.Phys.J)}}}%
\udk{539.172}
\nazvacol{A SIMULTANEOUS CENTER--OF--MASS CORRECTION}%
\avtor{A. SHEBEKO, P. GRYGOROV$^1$}%
\avtorcol{A. SHEBEKO, P. GRYGOROV}%
\inst{NSC "Kharkiv Institute of Physics and Technology"}%
\adr{(1, Akademicheskaya Str., Kharkiv 61108, Ukraine; e-mail: shebeko@kipt.kharkov.ua),}%
\insti{{}Department of Physics and
Technology, V.N. Karazin Kharkiv National University}%
\adri{(31, Kurchatova Ave., Kharkiv 61108, Ukraine; e-mail: pavel\_grigorov@mail.ru)}%

\begin{document}

\setcounter{page}{1}%
\maketitl                 
\begin{multicols}{2}
\anot{%
The approach exposed in the recent paper (A. Shebeko, P.
Papakonstantinou, E. Mavrommatis, Eur. Phys. J. A $\textbf{27}$,
$143$ ($2006$)) has been applied in studying center-of-mass motion
effects on the nucleon density and momentum distributions in
nuclei. We use and develop the formalism based upon the Cartesian
or boson representation, in which the coordinate and momentum
operators are expressed through the creation and annihilation
operators for oscillator quanta in the three different space
directions. We are focused upon effects due to the center-of-mass
and short-range nucleon correlations embedded in translationally
invariant ground-state wavefunctions. The latter are constructed
in the so-called fixed center-of-mass approximation, starting with
a Slater determinant wave function modified by some correlator
(\emph{e.g.}, after Jastrow or Villars). It is shown how one can
simplify evaluation of the corresponding expectation values that
determine the distributions. The analytic expressions derived here
involve the own "Tassie-Barker" \, factors for each distribution.
As an illustration, numerical calculations have been carried out
for the nucleus $^{4}He$ with the Slater determinant to describe
the nucleon $(1s)^4$ configuration composed of single-particle
orbitals which differ from  harmonic oscillator ones at small
distances. Such orbitals simulate somewhat short-range repulsion
between nucleons. Special attention is paid to a simultaneous
shrinking of the center--of--mass corrected density and momentum
distributions compared to the purely $(1s)^4$ shell
nontranslationally invariant
ones.
}%

\section{Introduction}

Treatment of the center--of--mass (CM) motion has been an
attractive subject of exploration in earlier and more recent
studies of nuclear theory (see, {\it e.g.},
\cite{Fria71}--\cite{ShePaMav06}). Those studies originated from
the necessity to remedy a deficiency of the nuclear many-body wave
function (WF), namely its lack of translational invariance (TI)
wherever shell-model single-particle (s.p.) WF's are used for its
construction. This deficiency is important in quite a number of
cases.

In the present investigation we adopt the "fixed-CM approximation"
\cite{RKB70,EST73} as a recipe to restore TI of a many-body WF
which does not possess this property. We apply it when evaluating
the elastic form factor (FF) $F(q)$ and the nucleon momentum
distribution (MD) $\eta(p)$ for light nuclei, and more
specifically for $^{4}He$ in its ground state (g.s.). Following
\cite{ShePaMav06} we prefer to deal with the intrinsic quantities
which are determined as expectation values of appropriate
(multiplicative) operators that depend on the corresponding Jacobi
variables and act on the intrinsic WF's. We have seen in
\cite{ShePaMav06} that the intrinsic density distribution (DD)
$\rho_{int}(r)$, being defined by the Fourier transform of $F(q)$,
does not coincide with the diagonal part of the one-body density
matrix (1DM), which is related in a standard manner to the
intrinsic MD. In the context, we also note that the term
"one-body" \, used here is somewhat conventional.
Let us mention that $F(q)$ and $\eta(p)$ can be related to the
different quantities measured via electron--nucleus collisions,
respectively, the elastic electron scattering cross sections and
the inclusive electron scattering cross sections. First of all, we
mean comparatively simple relations in the Born approximation with
the plane electron waves. In addition, to the so--called
approximation of small interaction times (see
\cite{DOS76}--\cite{Dem88} and refs. therein)
the double differential $(e,e')$ reaction cross section becomes
proportional to an integral of $\eta(p) $ over the momentum range
that is fixed with certain combination (the so-called $y$ -
scaling variable) of the momentum transfer $q$ and the energy
transfer $\omega $ (cf. \cite{Ciofi85}). Of course, in the
framework of these approximations one neglects off--shell effects
in the electron scattering on bound nucleons and  meson exchange
currents (MEC) contributions to an effective electromagnetic
(e.m.) interaction with nuclei. The latter should be taken into
account (see,\emph{ e.g.}, refs. \cite{KatAkaTan82,SchiPaWi90})
when describing the electron scattering on nuclei, especially at
high momentum transfers (in particular, helping to remove certain
discrepancy between theory and experiment in the vicinity of the
first minimum $|F_{ch}(q)|$ at $q^2 = 10 fm^{-2}$ for $^{4}He$).
Therefore, any comparison with experimental data omitting such
physical inputs has a restricted character. Nevertheless, in case
of light nuclei every approximate evaluation of intrinsic
quantities, being independent of different constraints originated
from reaction mechanisms, can be compared with microscopic
("exact") results. In the respect, our addressing to the alpha
particle seems to be perfectly explicable.

The aim of this paper is to show to what extent the approach
developed in refs. \cite{GonShe74,DOS76,KorShe77,ShePaMav06} can
be useful in calculations with more realistic WF's than simple
harmonic oscillator model (HOM) ones. In this connection, we
consider the CM correction of $F(q)$ and $\eta (p)$ treated on an
equal physical footing, {\it viz.}, by using one and the same
translationally invariant g.s. WF that incorporates the
nucleon-nucleon short-range correlations (SRC). One should note
that despite much interest over the last two decades concerning
the MD in nuclei \cite{FruMo84}--\cite{YpGr95},
its CM correction does not appear to have been properly treated
except in certain studies, where harmonic oscillator (HO) wave
functions were used (see, {\it e.g.}, \cite{DOS76}). Note also
calculations beyond HOM in \cite{GKSY02}. The underlying formalism
with basic definitions is exposed in the following section.
Section 3 contains analytic results of our derivations beyond HOM,
while the corresponding numerical results are discussed and
compared with experimental data in section 4.

\section{The intrinsic form factor, density and momentum distributions
with short-range correlations included}

By definition, the intrinsic (elastic) FF of a nonrelativistic
system with the mass number $A$ and the total angular momentum
equal to zero is
\begin{equation}
F(q)=F_{int} (q) \equiv  \langle \Phi_{int} \mid \exp[\imath \vec
{q} \cdot (\hat {\vec {r}}_1 - \hat {\vec {R}})] \mid \Phi_{int}
\rangle,
\end{equation}
where $\Phi_{int} $ is the intrinsic WF of the system (nucleus),
$\hat {\vec {r}}_1$ the coordinate operator for nucleon number 1,
and $\hat {\vec {R}} = A^{-1} \sum_{i=1}^A \hat {\vec {r}}_i$ the
CM operator.

In the fixed-CM approximation, according to the Ernst, Shakin and
Thaler (EST) prescription \cite{EST73} the nuclear many-body WF
with the total momentum $\vec{P}$ can be written in the form:
\begin{equation}
\mid \Psi_{P} \rangle =|\vec{ P}) \mid \Phi^{EST}_{int} \rangle,
\end{equation}
where a round bracket is used to represent a vector in the space
of the CM coordinate,\,so that $|\vec {P})$ means the eigenstate
of total momentum  operator $\hat{\vec {P}}$. The intrinsic WF
after EST
\begin{equation}
\mid \Phi^{EST}_{int} \rangle = \frac{ (\vec{R}=0 \mid \Phi
\rangle} {[\langle \Phi \mid \vec{R}=0)(\vec{R}=0 \mid \Phi
\rangle]^{1/2}}
\end{equation}
is constructed from an arbitrary (in general, translationally
non-invariant) WF $\Phi$, by requiring that the CM coordinate
$\vec {R}$ be equal to zero. The corresponding FF is the ratio
\[
F_{EST}(q) = \frac{A(q)}{A(0)},
\]
\begin{equation}
A(q)= \langle \Phi \mid {(2\pi)}^3 \delta (\hat {\vec {R}}) \exp[i
\vec {q} \cdot (\hat {\vec {r}}_1 - \hat {\vec {R}})] \mid \Phi
\rangle.
\end{equation}

Using the Cartesian representation in which
\begin{equation}
\hat{\vec{r}}=\frac{r_0}{\sqrt{2}}~(\hat{\vec{a}}^{\dag}+\hat{\vec{a}}),
~\hat{\vec{p}}=\imath\frac{p_0}{\sqrt{2}}~(\hat{\vec{a}}^{\dag}-\hat{\vec{a}})
\end{equation}
with the Bose commutation rules,
\begin{equation}
[\hat{a}_i,\hat{a}_j] = 0, \qquad
[\hat{a}_i,\hat{a}_j^{\dag}]=\delta_{ij}~~(i,j=1,2,3)
\end{equation}
and arbitrary real c-numbers $r_0$ and $p_0$ that meet the
condition
\begin{equation}
r_0p_0=1,
\end{equation}
one can show (see \cite{ShePaMav06}, \cite{KorShe77} and Appendix
{\rm A}) that
\begin{equation}
A(q)=\exp{\left(-\frac{\bar{r}_0^2 q^{2}}{4} \right)}U(q),
\end{equation}
\begin{equation}
U(q)=\int{d\vec{\lambda}\exp{\left(-\frac{r_0^2\lambda^2}{4A}\right)}
F(\vec{v},\vec{s})},
\end{equation}
\begin{equation}
F(\vec{v},\vec{s})=\langle\Phi|\hat{O}_1(\vec{v}+\vec{s}~)\hat{O}_2(\vec{v})\dots\hat{O}_A(\vec{v})|\Phi\rangle,
\end{equation}
where
\begin{equation}
\hat{O}_\gamma(\vec{x})=\exp(-\vec{x}^\ast\hat{\vec{a}}^\dag_\gamma)\exp(\vec{x}\hat{\vec{a}}_\gamma)
\equiv\hat{E}^\dag_\gamma(-\vec{x})\hat{E}_\gamma(\vec{x})
\end{equation}
\[
(\gamma=1,\dots,A)
\]
with
\begin{equation}
\vec{s}=\imath\frac{r_0}{\sqrt{2}}~\vec{q},~\vec{v}=\imath\frac{r_0}{\sqrt{2}A}(\vec{\lambda}-\vec{q})
\end{equation}
and the renormalized "length" parameter
\[
\bar{r}_0 = \sqrt{\frac{A-1}{A}}r_0.
\]

Further, starting from the definition of the intrinsic MD (see
\cite{ShePaMav06}),
\begin{equation}
\eta(p) = \langle \Phi_{int} \mid \delta (\hat {\vec {p}}_1 - \hat
{\vec {P}}/A - \vec{p}) \mid \Phi_{int} \rangle,
\end{equation}
we consider the distribution in the fixed-CM approximation:
\begin{equation}
\eta_{EST}(p) = \frac{\langle \Phi \mid {(2\pi)}^3 \delta (\hat
{\vec {R}}) \delta (\hat {\vec {p}}_1 - \hat {\vec {P}}/A -
\vec{p}) \mid \Phi \rangle}{\langle \Phi \mid {(2\pi)}^3 \delta
(\hat {\vec {R}}) \mid \Phi \rangle}
\end{equation}
and the Fourier transform
\begin{equation}
\eta_{EST}(p)=(2\pi)^{-3}\int\exp(-\imath\vec{p}\vec{x})N(x)/N(0)d\vec{x}
\end{equation}
with
\begin{equation}
N(x)=\langle \Phi\mid (2\pi)^{3}
\delta(\vec{R})\exp[\imath(\vec{p}_1-\vec{P}/A)\vec{x}]\mid\Phi\rangle.
\end{equation}
We see the certain resemblance between the structure functions
$N(x)$ and $A(q)$, \emph{viz.}, both are determined by the
expectation values of similar multiplicative operators with one
and the same trial WF $\Phi$. Owing to this, using the same
algebraic technique we get
\begin{equation}
N(x)=\exp\left( - \frac{\bar{p}_0^2 x^2}{4}  \right) D(x),
\end{equation}
\begin{equation}
D(x)=\int d \vec{\lambda} \exp\left(-\frac{r_0^2\lambda^2}{4A}
\right)F(\vec{v}~',\vec{s}~'),
\end{equation}
where
\begin{equation}
\vec{s}~'=-\frac{p_0}{\sqrt{2}}\vec{x},~~~\vec{v}~'=\frac{\imath
r_0}{\sqrt{2}A}(\vec{\lambda}-\imath p_0^2\vec{x})
\end{equation}
and
\[
\bar{p}_0 = \sqrt{\frac{A-1}{A}}p_0.
\]
Certain relation of the MD to the corresponding intrinsic density
matrix has been shown in \cite{ShePaMav06}.

After this let us assume a trial WF,
\begin{equation}
\mid \Phi \rangle=\mid \Phi_{corr} \rangle =\hat{C}(1,2, \cdots, A
)\mid Det \rangle
\end{equation}
with the Slater determinant
\begin{equation}
\mid Det \rangle=\frac{1}{\sqrt{A!}}\sum_{\it \hat{\mathcal{P}}
\in S_A} \epsilon_{ \mathcal{P}} \hat{\mathcal{P}}\{\mid
\phi_{p_1}(1)\rangle \dots \mid \phi_{p_A}(A)\rangle\}.
\end{equation}
Here $\epsilon_{ \mathcal{P}}$ is the parity factor for the
permutation $\mathcal{P}$, $\phi_a$ the occupied orbital with the
quantum numbers $\{a\}$ and the summation runs over all
permutations of the symmetric group $S_A$.

The \emph{A}-particle operator $\hat{C} =
\hat{C}(\hat{\vec{r}}_\alpha-\hat{\vec{r}}_\beta,
~\hat{\vec{p}}_\alpha-\hat{\vec{p}}_\beta) $ \footnote{Of course,
the operator may be spin and isospin dependent} introduces the SRC
and meets all necessary requirements of the translational and
Galileo invariance, the permutable and rotational symmetry, etc.
However, being translationally invariant itself such a model
introduction of correlations does not enable to restore the TI
violated with such a shell-model WF as the Slater determinant.

What it follows can be used with the Jastrow correlator
\cite{Jast55}
\begin{equation}
\hat{C}= \prod\limits_{\alpha < \beta}^A
f(\hat{\vec{r}}_{\alpha\beta} ),
\end{equation}
where $f(\hat{\vec{r}}_{\alpha\beta} )$  is a two-body correlation
factor whose deviation from unity occurs only for small distances
$r_{\alpha\beta} = |{\vec{r}}_\alpha - {\vec{r}}_\beta|$ less than
a correlation radius $r_c$.

Another popular option goes back to the lectures by Villars in
\cite{Villa63} (see also \cite{ProviShak64}) with a unitary
operator
\begin{equation}
\hat{C}=\exp(-\imath \hat{G}),
\end{equation}
\begin{equation}
\hat{G}=\sum\limits_{\alpha<\beta}\hat{g}(\alpha,\beta),
\end{equation}
where the Hermitian operator $\hat{g}(\alpha,\beta)$ acts onto the
space of the pair $(\alpha,\beta)$. In particular, we could follow
the simplest Darmstadt ansatz \cite{Feld98}:
\begin{equation}
\hat{g}(\alpha,\beta)
=\frac{1}{2}\{\vec{s}~(\hat{\vec{r}}_{\alpha\beta})\hat{\vec{p}}_{\alpha\beta}+
\hat{\vec{p}}_{\alpha\beta}\vec{s}~(\hat{\vec{r}}_{\alpha\beta})
\},
\end{equation}
where $\vec{s}$ is a function of the relative coordinate
$\hat{\vec{r}}_{\alpha\beta}=\hat{\vec{r}}_\alpha-\hat{\vec{r}}_\beta$.
Its canonically conjugate momentum
$\hat{\vec{p}}_{\alpha\beta}=\frac{1}{2}(\hat{\vec{p}}_\alpha-\hat{\vec{p}}_\beta)$.

Keeping in mind similar constructions we rewrite expectation (10)
as
\begin{equation}
F(\vec{v},\vec{s})=\langle\Phi(-\vec{v})\mid
\hat{E}^\dag_1(-\vec{s})\hat{E}_1(\vec{s})\mid\Phi(\vec{v})\rangle,
\end{equation}
where
\[
\mid\Phi(\vec{x})\rangle=\hat{E}_1(\vec{x})\dots\hat{E}_A(\vec{x})\mid\Phi\rangle,
\]
since
$\hat{E}_1(\vec{v}+\vec{s})=\hat{E}_1(\vec{v})\hat{E}_1(\vec{s})$
and
$[\hat{E}_\alpha(\vec{x}),\hat{E}_\beta(\vec{y})]=0~~\\
(\alpha,\beta=1,\dots,A)$ for any vectors $\vec{x}$ and $\vec{y}$.

Moreover, we find that
\begin{equation}
\hat{E}(\vec{x})~\hat{\vec{r}}~\hat{E}^{-1}(\vec{x})=\hat{\vec{r}}+\frac{r_0}{\sqrt{2}}~\vec{x}
\end{equation}
and
\begin{equation}
\hat{E}(\vec{x})\hat{\vec{p}}~\hat{E}^{-1}(\vec{x})=\hat{\vec{p}}-\imath\frac{p_0}{\sqrt{2}}~\vec{x}.
\end{equation}
Remind that $E^\dag\neq E^{-1}$. In other words,
$\hat{E}_\alpha(\vec{x})$ is the displacement operator in the
space of nucleon states with the label $\alpha$.

Due to this property when handling the similarity transformation
\[
\hat{C}'=\hat{E}_1(\vec{x})\dots\hat{E}_A(\vec{x})
\hat{C}(\hat{\vec{r}}_\alpha-\hat{\vec{r}}_\beta,~\hat{\vec{p}}_\alpha-
\hat{\vec{p}}_\beta)\times
\]
\[
\times\hat{E}^{-1}_1(\vec{x})\dots\hat{E}^{-1}_A(\vec{x}),
\]
we get
\[
\hat{C}'=\hat{C}(\hat{E}_\alpha(\vec{x})\hat{\vec{r}}_\alpha\hat{E}^{-1}_\alpha(\vec{x})-
\hat{E}_\beta(\vec{x})\hat{\vec{r}}_\beta\hat{E}^{-1}_\beta(\vec{x}),
\]
\[
\hat{E}_\alpha(\vec{x})\hat{\vec{p}}_\alpha\hat{E}^{-1}_\alpha(\vec{x})-
\hat{E}_\beta(\vec{x})\hat{\vec{p}}_\beta\hat{E}^{-1}_\beta(\vec{x}))=
\]
\[
=\hat{C}(\vec{r}_\alpha-\vec{r}_\beta,\vec{p}_\alpha-\vec{p}_\beta)=\hat{C}
\]
\emph{i.e.},
\begin{equation}
\hat{C}'=\hat{C}.
\end{equation}
Recall that $C$ is a function of \underline{all} the relative
coordinates and their
canonically conjugate momenta.\\
From eqs. (20) and (29) it follows that
\[
\mid \Phi_{corr}(\vec{x})\rangle \equiv
\hat{E}_1(\vec{x})\dots\hat{E}_A(\vec{x}) \mid \Phi_{corr}
\rangle=
\]
\begin{equation}
=\hat{C}\mid Det(\vec{x})\rangle.
\end{equation}
Here $\mid
Det(\vec{x})\rangle= \hat{E}_1(\vec{x})\dots\hat{E}_A(\vec{x})\mid
Det \rangle$ is a new Slater determinant composed of the
renormalized orbitals,
\begin{equation}
\mid \phi_{a}(\vec{x};\alpha)\rangle=
\hat{E}_{\alpha}(\vec{x})\mid\phi_{a}(\alpha)\rangle~~~(\alpha=1,\dots,A),
\end{equation}
\emph{viz.},
\begin{equation}
\mid Det (\vec{x}) \rangle=\frac{1}{\sqrt{A!}}\sum_{\it
\hat{\mathcal{P}} \in S_A} \epsilon_{ \mathcal{P}}
\hat{\mathcal{P}}\{\mid \phi_{p_1}(\vec{x};1)\rangle \dots \mid
\phi_{p_A}(\vec{x};A)\rangle\}.
\end{equation}
In their turn, such orbitals can be evaluated in a concise
analytic form as initial ones are linear combinations of the HOM
orbitals (see Appendix A).

Following (26) we arrive to
\[
F_{corr}(\vec{v},\vec{s}) \equiv \langle\Phi_{corr}(-\vec{v})\mid
\hat{E}_1^\dag(-\vec{s})\hat{E}_1(\vec{s})\mid\Phi_{corr}(\vec{v})\rangle=
\]
\begin{equation}
=\langle Det(-\vec{v})\mid
\hat{C}^\dag\hat{E}_1^\dag(-\vec{s})\hat{E}_1(\vec{s})\hat{C} \mid
Det(\vec{v})\rangle.
\end{equation}
Expressions (8) and (17) with expectations $F(\vec{v},\vec{s})$
and $F(\vec{v}~',\vec{s}~')$, which are determined by eq. (33),
are certain base for our calculations.

\subsection{Several working formulae: application to $^{4}He$}

In special case of the pure HOM $(1s)^4$ configuration occupied by
the four nucleons in $^4 He$ we have
\begin{equation}
\mid \Phi_{corr} (\vec{x})\rangle = \mid \Phi_{corr} \rangle=
\hat{C} \mid (1s)^4 \rangle,
\end{equation}
taking into account that the HOM g.s. $\mid (1s)^4\rangle$ is the
vacuum for operators $\hat{\vec{a}}_\alpha$
($\alpha=1$,\dots,$A$). It is the case, where $\mid
Det(\vec{v})\rangle$ does not depend on $\vec{v}$ coinciding with
the initial Slater determinant $\mid (1s)^4 \rangle$. Hence,
\begin{equation}
F_{1s}(\vec{v},\vec{s})=\langle (1s)^4 \mid
\hat{C}^\dag\hat{E}_1^\dag(-\vec{s})\hat{E}_1(\vec{s})\hat{C} \mid
(1s)^4 \rangle.
\end{equation}
In other words, under such a simplification the function
$F(\vec{v},\vec{s})$ in integral (9) becomes independent of
$\vec{\lambda}$ and we get
\[
U(q)=U_{1s}(q)\int \exp\left(-\frac{r_0^2\lambda^2}{4A}\right)
d\vec{\lambda},
\]
so that
\begin{equation}
\frac{U_{1s}(q)}{U_{1s}(0)}=\frac{\langle (1s)^4 \mid
\hat{C}^\dag\hat{E}_1^\dag(-\imath\frac{r_0}{\sqrt{2}}\vec{q})\hat{E}_1(\imath\frac{r_0}{\sqrt{2}}\vec{q})\hat{C}
\mid (1s)^4 \rangle}{\langle (1s)^4 \mid \hat{C}^\dag \hat{C} \mid
(1s)^4 \rangle}.
\end{equation}
Thus, the FF of interest is
\begin{equation}
F_{EST}(q)=F_{TB}(q)F_{IPM}(q)F_{corr}(q),
\end{equation}
where according to eq.(A.1) we have the Tassie-Barker $F_{TB}(q)$
and the HOM FF $F_{HOM}(q)$. The factor
\begin{equation}
F_{corr}(q)=\frac{\langle (1s)^4 \mid
\hat{C}^\dag\hat{E}_1^\dag(-\imath\frac{r_0}{\sqrt{2}}\vec{q})
\hat{E}_1(\imath\frac{r_0}{\sqrt{2}}\vec{q})\hat{C}\mid (1s)^4
\rangle} {\langle (1s)^4 \mid \hat{C}^\dag \hat{C} \mid (1s)^4
\rangle}
\end{equation}
incorporates the SRC in any way.

At this point, one can proceed, at least, along the two
guidelines. One of them could be based upon the representation
\[
\langle (1s)^4 \mid
\hat{C}^\dag\hat{E}_1^\dag(-\imath\frac{r_0}{\sqrt{2}}\vec{q})
\hat{E}_1(\imath\frac{r_0}{\sqrt{2}}\vec{q})\hat{C} \mid (1s)^4
\rangle=
\]
\begin{equation}
=\langle (1s)^4 \mid \hat{C}^\dag_1 (-\vec{q})\hat{C}_1(\vec{q})
\mid (1s)^4 \rangle,
\end{equation}
where
\[
\hat{C}_1(\vec{q})=\hat{E}_1(\imath\frac{r_0}{\sqrt{2}}\vec{q})
\hat{C}(\hat{\vec{r}}_1, \hat{\vec{p}}_1, \cdots)
\hat{E}_1^{-1}(\imath\frac{r_0}{\sqrt{2}}\vec{q})= \
\]
\begin{equation}
=C\left(\hat{\vec{r}}_1+\imath\frac{\vec{q}}{2}r_0^2,
\hat{\vec{p}}_1 + \frac{\vec{q}}{2}, \dots\right).
\end{equation}
Other continuation is prompted by the relation
\[
\hat{E}^\dag_1(-\imath\frac{r_0}{\sqrt{2}}~\vec{q}~)\hat{E}_1(\imath\frac{r_0}
{\sqrt{2}}~\vec{q}~)=\exp\left(\frac{r_0^2q^2}{4}\right)\exp(\imath\vec{q}~\hat{\vec{r}}_1),
\]
that gives rise to
\begin{equation}
F_{corr}(q)=\exp\left( \frac{r_0^2 q^2}{4}\right) F_C (q),
\end{equation}
\begin{equation}
F_C(q)=\frac{\langle (1s)^4\mid \hat{C}^\dag \exp(\imath
\vec{q}\hat{\vec{r}}_1)\hat{C}\mid (1s)^4 \rangle}{\langle
(1s)^4\mid \hat{C}^\dag \hat{C}\mid (1s)^4 \rangle},
\end{equation}
where $F_C(q)$ is the no CM corrected FF with the correlated g.s.
$\hat{C}\mid(1s)^4\rangle$.

Analogously, we find
\begin{equation}
N(x)=N_{TB}(x)\,N_{HOM}(x)\,N_{corr}(x)
\end{equation}
with the \underline{own} Tassie-Barker factor
\begin{equation}
N_{TB}(x)=\exp\left(\frac{p_0^2~x^2}{4A} \right)
\end{equation}
and
\begin{equation}
N_{HOM}(x)=\exp\left(-\frac{p_0^2~x^2}{4} \right),
\end{equation}
\begin{equation}
N_{corr}(x)=\frac{\langle (1s)^4\mid
\hat{C}^\dag\hat{E}^\dag_{1}(\frac{p_0}{\sqrt{2}}\vec{x})\hat{E}_{1}(-\frac{p_0}{\sqrt{2}}\vec{x})\hat{C}\mid
(1s)^4 \rangle}{\langle (1s)^4\mid \hat{C}^\dag\hat{C}\mid (1s)^4
\rangle}.
\end{equation}

Again, different continuations are possible (cf. the transition
from eq. (38) to eqs. (39) and (41)). In particular, with the help
of
\[
\hat{E}^\dag_{1}(\frac{p_0}{\sqrt{2}}\vec{x})\hat{E}_{1}(-\frac{p_0}{\sqrt{2}}\vec{x})=
\exp\left(\frac{p_0^2~x^2}{4}\right)\exp\left(\imath
\hat{\vec{p}}_1\vec{x}\right)
\]
we get
\begin{equation}
N_{corr}(x)=\exp\left(\frac{p_0^2~x^2}{4}\right)N_C(x),
\end{equation}
\begin{equation}
N_{C}(x)=\frac{\langle (1s)^4\mid
\hat{C}^\dag\exp(\imath\hat{\vec{p}}_1\vec{x})\hat{C}\mid (1s)^4
\rangle}{\langle (1s)^4\mid \hat{C}^\dag\hat{C}\mid (1s)^4
\rangle}.
\end{equation}
The Fourier transform
\begin{equation}
\eta_C(p)=\frac{1}{(2\pi)^3}\int {\rm
e}^{-\imath\vec{p}\vec{x}}N_C(x) d\vec{x}
\end{equation}
gives us the one-body momentum distribution (OBMD) without the CM
correction of the model g.s. $\hat{C}\mid(1s)^4\rangle$.

By definition, the intrinsic DD is
\[
\rho_{int}(r)=\frac{1}{(2\pi)^3}\int {\rm
e}^{-\imath\vec{q}\vec{r}}F_{int}(q) d\vec{q} =
\]
\begin{equation}
=\langle \Phi_{int} \mid \delta (\hat {\vec {r}}_1 - \hat {\vec
{R}} - \vec{r}) \mid \Phi_{int} \rangle,
\end{equation}
so that the relations
\[
\rho_{EST}(r)=\frac{1}{(2\pi)^3}\int {\rm
e}^{-\imath\vec{q}\vec{r}}F_{EST}(q) d\vec{q}
\]
and
\[
\rho_{C}(r)=\frac{1}{(2\pi)^3}\int {\rm
e}^{-\imath\vec{q}\vec{r}}F_{C}(q) d\vec{q}
\]
are, respectively, the one-body density distribution (OBDD) with
the CM correction and the no CM corrected distribution.

Thus, we have shown (with the help of purely algebraic means) that
evaluation of the distributions can be reduced to the well-known
treatment. Indeed, expectations values (42) and (48) occur in all
conventional calculations with the many-particle WF (20),
\emph{i.e.,} without any CM correction. Diverse methods have been
elaborated when evaluating similar quantities (see, \emph{e.g.,}
 \cite{MihHeis99}, \cite{Navrat04}, \cite{ProviShak64},
 \cite{Feld98}, \cite{GauGilRip1971},
\cite{BohStrin82},
 \cite{Moust01},
\cite{Alviol05} and refs. therein). In this work we confine
ourselves to comparatively simple computations for a $(1s)^4$
configuration, where a short-range repulsion between nucleons is
introduced in an effective way, \emph{viz.,} modifying the s.p.
orbital as in \cite{RKB70}. Respectively, the following WF is used
in the next section.

\section{Analytic expressions for the form factor, density and momentum
 distributions with the single-particle wave function beyond HOM}

In accordance with \cite{RKB70} we employ the normalized
Radhakant, Khadkikar and Banerjee (RKB) radial orbital for the
lowest s.p. state of $^4He$,
\begin{equation}
\phi^{RKB}(r)=\frac{1}{\sqrt{1+\beta ^2}}(\phi_{00}(r)+\beta
\phi_{10}(r)),
\end{equation}
 where $\phi_{00}$ and $\phi_{10}$ are the
normalized HO radial eigenfunctions:
\begin{equation}
\phi_{00}(r)=2\sqrt{{{1}\over{\sqrt{\pi} b_{H}}}}\frac{r}{b_H}
\exp\left(- \frac{r^2}{2b_{H}^2}\right),
\end{equation}

\begin{equation}
\phi_{10}(r)=\sqrt{{{3!}\over{\sqrt{\pi} b_{H}}}} \frac{r}{b_H}
\left[1-{{2}\over{3}}\frac{r^2}{b_{H}^2}\right] \exp\left(-
\frac{r^2}{2b_{H}^2}\right)
\end{equation}
for the states with $n=0, l=0$ and $n=1, l=0$, respectively. Here
$b_H$ is the HO parameter and $\beta$ is a mixing parameter.

The RKB WF allows one to obtain the following expressions for the
density distribution (normalized to unity), for the point proton
FF as well as for the MD (also normalized to unity):
\[
{\rho}_{sp}^{RKB}(r)=\frac{1}{(\sqrt{\pi} b_{H})^{3}
(1+{\beta}^{2})} \exp\left(- \frac{r^2}{b_{H}^2}\right)\times
\]
\begin{equation}
\times~\left[1+\sqrt{\frac32} \beta \left(1-\frac{2r^2}{3
b_H^2}\right) \right]^2 ,
\end{equation}
\[
F_{sp}^{RKB}(q)=\frac{1}{1+{\beta}^{2}}\exp\left(-\frac{(b_{H}q)^2}{4}\right)\times
\]
\begin{equation}
\times\left[ 1+{\beta}^{2}+\frac{\beta}{\sqrt{6}} \left(1 -
\sqrt{\frac23} \beta \right) b_{H}^2 q^{2} + \frac{\beta^2 b_{H}^4
q^{4}}{24} \right],
\end{equation}
\[
\eta_{sp}^{RKB}(p)=\frac{b_H^3}{\pi\sqrt{\pi}(1+\beta^2)}\exp(-b_H^2
p^2)\times
\]
\begin{equation}
\times~\left[1 - \sqrt{\frac32} \beta \left(1-\frac23b_H^2
p^2\right)\right]^2.
\end{equation}

\subsection{ The CM corrected form factor $F(q)$ and its reduction
to quadratures}

Assuming a Slater determinant as the g.s. $(1s)^4$ of $^4He$ its
FF in the fixed-CM approximation can be written in the form (cf.
\cite{RKB70}):
\begin{equation}
F_{EST}(q) = {{ \int F_{1s}(|\vec{q}+\vec{u}|) F_{1s}^3(u)
d\vec{u} }\over { \int F_{1s}^4(u) d\vec{u} }},
\end{equation}
where
\[
F_{1s}(v) \equiv \int {\rm e}^{i\vec{v} \vec{r} }\phi _{1s}^2 (r)
{\rm d}\vec{r}={ {4\pi}\over {v} } \int \phi _{1s}^2 (r) \sin vr \
r{\rm d}r
\]
is the no CM-corrected FF.

In case of the RKB--like s.p. WFs whose orbitals are truncated
expansions in the radial HO eigenfunctions the multiple integrals
in the r.h.s. of eq.(63) can be expressed through simple
integrals. The respective algebraic technique has been developed
in ref. \cite{DOS76} and exposed recently in \cite{She00} (see
also Appendix A to the present paper). Its application with the
RKB orbital enables us to get
\begin{equation}
F^{RKB}(q)=\frac{A^{RKB}(q)}{A^{RKB}(0)},
\end{equation}
where
\[
A^{RKB}(q)=I_{1}(q)+I_{2}(q),
\]
\[
I_{1}(q)={\frac{4\pi}{{qb_{H}^{4}}}}\exp\left(-\frac{3}{16} q^2
b_H^2\right)\times
\]
\[
\times~\int\limits_0^\infty \{B_{2}\left[\frac14(t-\frac34
~qb_{H})^{2}\right]
 M_2^3\left[\frac14 (t + \frac14~ qb_{H})^{2}\right]-
\]
\[
- B_2\left[\frac14(t + \frac34 ~qb_{H})^{2}\right]
M^3_2\left[\frac14(t - \frac14~ qb_{H})^{2}\right] \} \exp(- t^2)
t dt,
\]
\[
I_2(q)= {{\pi}\over{b_{H}^{3}}} \exp\left(-\frac{3}{16} q^2
b_H^2\right)\times
\]
\[
\times~\int\limits_0^\infty \{ B_{2}\left[\frac14 (t-\frac34
qb_{H})^{2}\right] M_2^3\left[\frac14 (t + \frac14
qb_{H})^{2}\right]+
\]
\[
+ B_2\left[\frac14(t + \frac34 qb_{H})^{2}\right]
M^3_2\left[\frac14(t - \frac14 qb_{H})^{2}\right] \} \exp(- t^2)
dt,
\]
\[
A^{RKB}(0)={{4\pi}\over{b_{H}^{3}}}\int\limits_0^\infty M_2^4
\left[\frac14 ~t^2\right]\exp(-t^2) t^2 dt.
\]
The functions $M_{2}(z)$ and $B_{2}(z)$ are second
degree polynomials of the variable $z$
\begin{center} \noindent
\epsfxsize=\columnwidth\epsffile{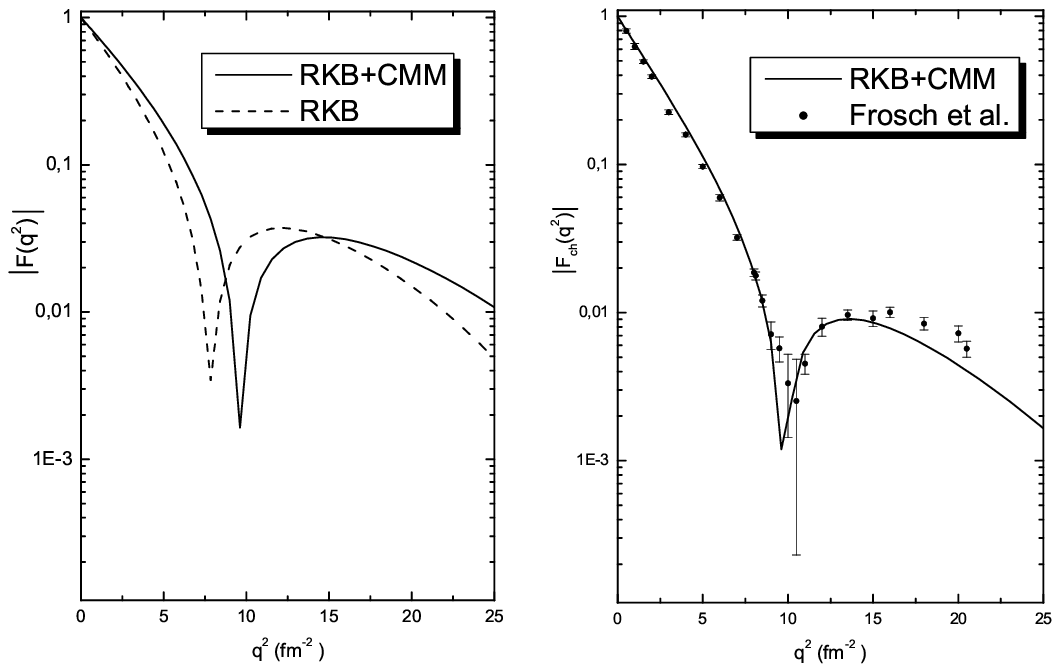}
\end{center}
\vskip-3mm\noindent{\footnotesize Fig.1. The point-like FF (left)
and the charge FF (right) of $^{4}He$. Curves calculated with RKB
WF using the EST prescription (solid) and without the
CM-correction (dashed); experimental points from \cite{frosh66}.
Other clarifications in the text
}%
\vskip15pt
\[
M_{2}(z)=m_{0}+m_{1}z+m_{2}z^{2}
\]
\[
B_{2}(z)=h_{0}+h_{1}z+h_{2}z^{2},
\]
where the constants are related to the mixing parameter $\beta$
\[
m_{0}=1+{\beta}^{2},
\]
\[
m_{1}=2\sqrt{2/3} \beta (1-\sqrt{2/3} \beta ),
\]
\[
m_{2}=(2/3){\beta}^{2},
\]
and
\[
h_{0}=m_{0}+m_{1}+2m_{2,}
\]
\[
h_{1}=m_{1}+2m_{2},
\]
\[
h_{2}=m_{2}.
\]

\begin{center} \noindent
\epsfxsize=\columnwidth\epsffile{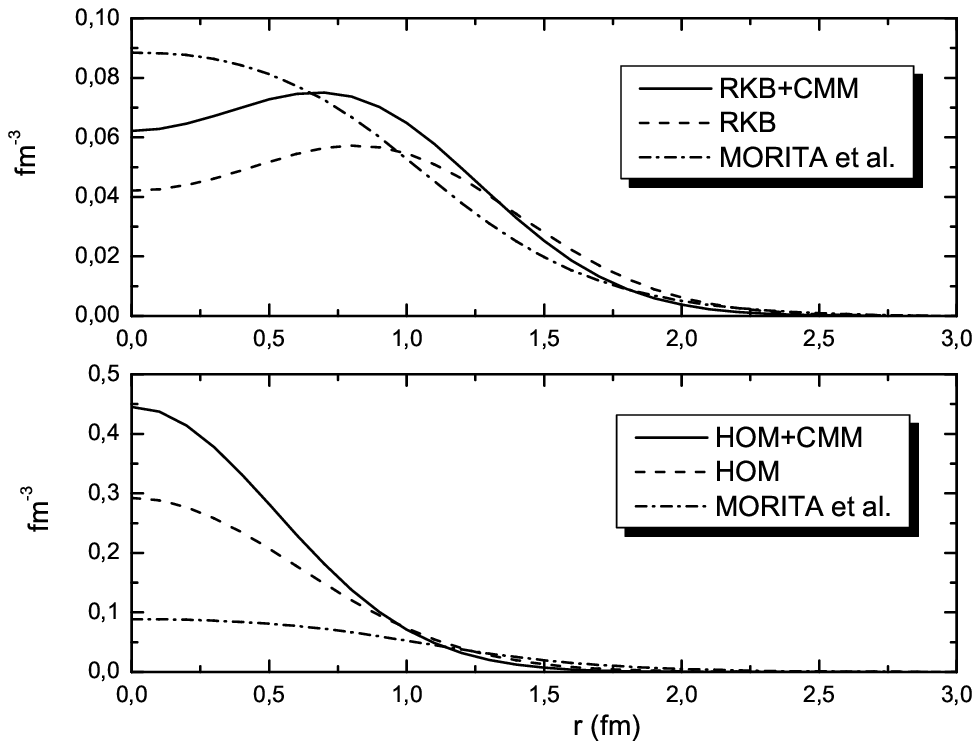}
\end{center}
\vskip-3mm\noindent{\footnotesize Fig.2. The one-body density
distribution (OBDD) for $(1s)^4$ configuration with RKB orbital in
the fixed-CM approximation (solid) and without the CM correction
(dashed). For two sets of parameters: $b_H=0.8532$ fm and
$\beta=-0.4738$ (top); $b_H=0.8532$ fm and $\beta=0$ (bottom).
Dot-dashed with the parametrization from \cite{Morita87}}
\vskip15pt

\subsection{The CM corrected momentum distribution $\eta(p)$ and
its reduction to quadratures}

In parallel, starting from eq.(16), we obtain for the $(1s)^4$
configuration with the Slater determinant $\mid \Phi \rangle =\mid
(1s)^4 \rangle $ (see Appendix ${\rm B}$ to Lect.I in
\cite{She00}):
\[
N_{EST}(x) = \int {\rm d}\vec {k}\ \langle 1s \mid
\exp\left(\imath\frac{\vec {k}\hat{\vec {r}}}{A}\right)
\exp\left(\imath \frac{A-1}{A} \hat{\vec {p}}\vec {x}\right ) \mid
1s \rangle\times
\]
\begin{equation}
\times \langle 1s \mid \exp\left(\imath\frac{\vec {k}\hat{\vec
{r}}}{A}\right) \exp\left(-\imath \frac{\hat{\vec {p}}\vec {x}}{A}
\right ) \mid 1s \rangle ^3.
\end{equation}

\begin{center} \noindent \epsfxsize=\columnwidth\epsffile{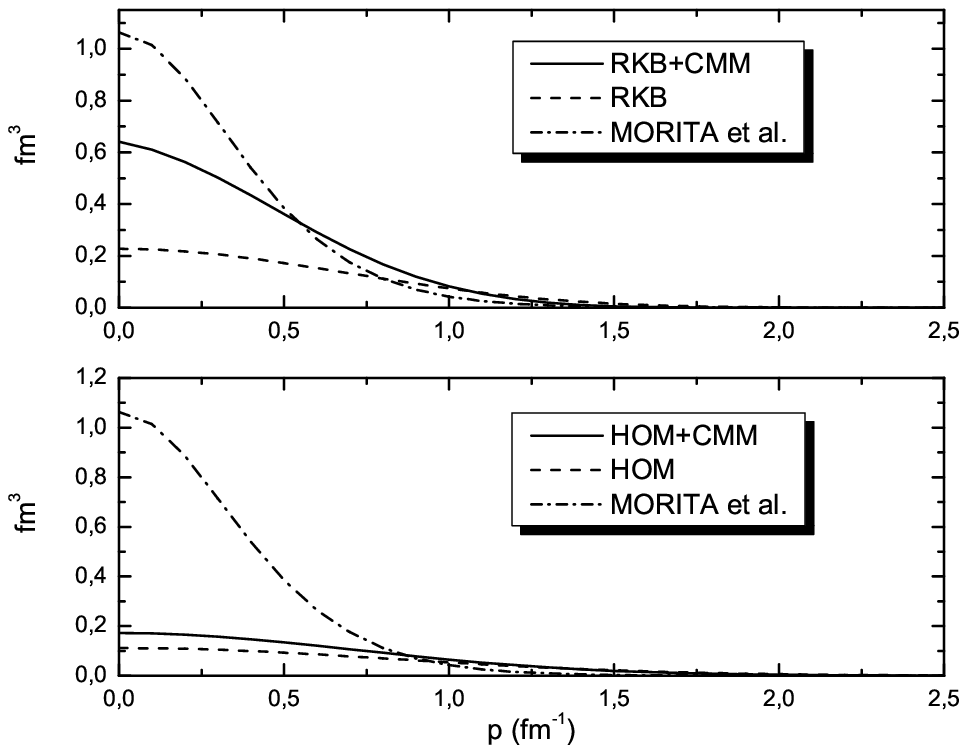}
\end{center}
\vskip-3mm\noindent{\footnotesize Fig.3. The one-body momentum
distribution (OBMD) for the $(1s)^4$ configuration with RKB
orbital. Difference between the curves is the same as in fig. 2.
Dot-dashed with parametrization from \cite{Morita88} } \vskip15pt

Again, using the representation (5) and splitting the exponents
involved in eq.(67) with the succesive normal ordering of the
operators $\hat{\vec{a}}~^\dag$ and $\hat{\vec{a}}$ (the former
are to the left from the latter), one can show that
\begin{equation}
{{N(x)^{RKB}}\over{N(0)^{RKB}}} =
\exp\left(-{{A-1}\over{A}}{{x^{2}}\over{4 b^{2}_{H}}}\right)
{{J(x)}\over{J(0)}},
\end{equation}
where the integral $J(x)$ is determined by
\[
J(x)=\int\limits^\infty_0\exp{\left(-\frac{r_0^2\lambda^2}{4A}\right)}g(\lambda^2;x^2)\lambda^2d\lambda,
\]
\[
g(\lambda^2;x^2)
=\left(A_1+\frac13A_2\right)B_1^3+\left(A_1+\frac35A_2\right)B_1^2B_2+
\]
\[
+3\left(\frac15 A_1+\frac17 A_2\right)B_1B_2^2+\left(\frac17
A_1+\frac19 A_2\right)B_2^3,
\]
\[
A_1=1+\sbt-\sqrt{\frac{2}{3}}~\bt\left[1+\sqrt{\frac{2}{3}}\bt\right]\left(\frac{A-1}{A}\right)^2\frac{\spo\sx}{2}+
\]
\[
+\sqrt{\frac{2}{3}}\bt\left[1-\sqrt{\frac{2}{3}}\bt\right]~\frac{\sro\sl}{2A^2}+
\]
\[
+\frac{1}{6}\sbt\left[\left(\frac{A-1}{A}\right)^2\frac{\spo\sx}{2}~-\frac{\sro\sl}{2A^2}\right]^2,
\]
\[
A_2=\frac{1}{6}\sbt\left(\frac{A-1}{A}\right)^2\frac{\sx\sl}{A^2},
\]
\[
B_1=1+\sbt-\sqrt{\frac{2}{3}}\bt\left[1+\sqrt{\frac{2}{3}}\bt\right]\frac{\spo\sx}{2A^2}+
\]
\[
+\sqrt{\frac{2}{3}}\bt\left[1-\sqrt\frac{2}{3}\bt\right]\frac{\sro\sl}{2A^2}+\frac{1}{6}\sbt\left[\frac{\spo\sx}{2A^2}-
\frac{\sro\sl}{2A^2}\right]^2,
\]
\[
B_2=\frac{1}{6}\sbt\frac{\sx\sl}{A^4}
\]
Thus, the structure function $N^{RKB}(x)$ can be reduced to
one-dimensional integrals similar to those derived for
$F^{RKB}(q)$. Here $A=4$, but we allow $A$ to be changeable,
particularly, in order to check that the corresponding
distribution
\[
\eta^{RKB}_{EST}(p)= \frac{1}{2\pi^2p}\int\limits_0^\infty
N^{RKB}(x)/N^{RKB}(0) \sin(px)xdx
\]
to the limit $A\rightarrow\infty$ yields the no CM corrected
distribution $(62)$.

\section{Results and discussion }

Analytic expressions obtained in sect.2 for the density and
momentum distributions and their Fourier transforms are
sufficiently general to be applied in different translationally
invariant treatments with the SRC included. The corresponding
formulae derived in sect.3 in case of the $^4He$ nucleus have been
employed to carry out our calculations beyond the simple HOM.
Their numerical results are displayed in figs.1-4.

\begin{center} \noindent \epsfxsize=\columnwidth\epsffile{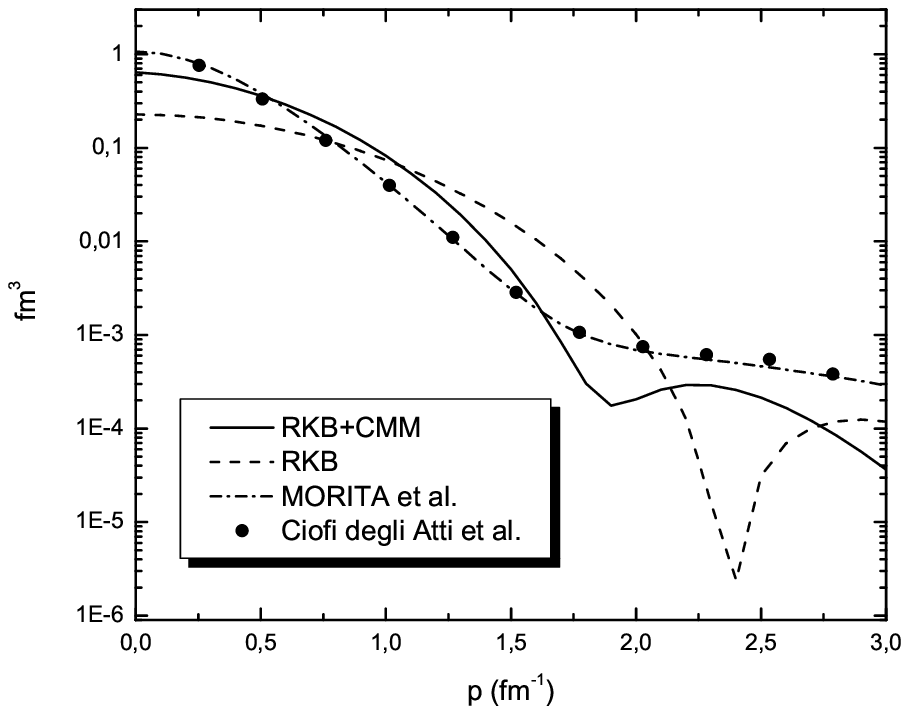}
\end{center}
\vskip-3mm\noindent{\footnotesize Fig.4. Variations
$log~\eta^{RKB}_{EST}(p)$ (solid), $log~\eta^{RKB}_{sp}(p)$
(dashed) and $log~\eta_{Morita}(p)$ (dot-dashed) with $p$.
Notation $log~\eta_{Morita}(p)=\left(\frac43\right)^3
W^{SN}(\frac43p)$, where function $W^{SN}(x)$ calculated with a
convenient parametrization from \cite{Morita88}. Points are
 resulted from \cite{Ciofi91} } \vskip15pt

In fig.\,1 we show our calculation of the charge FF, $F_{ch}(q^2)
= f_{p}(q)F(q) $ of the alpha particle, using eq.($55$) and
eq.($58$) and considering for the finite proton size factor
$f_{p}(q)$ the Chandra and Sauer prescription \cite{ChanSau}. The
two parameters $b_{H}$ and $\beta$ have been determined by the
least-square fitting to the experimental values \cite{frosh66}:
their best-fit values are $b_{H}=0.8532$ fm and $\beta=-0.4738$
($\chi^2 \simeq 13.07$). These values have been utilized in our
calculations shown in figs.\,2-4.

As it is seen in fig.1 (its left part), the CM correction leads to
a considerable qualitative change of $q$-dependence of the FF: its
first minimum and second maximum are shifted towards higher
$q$-values. This difference between solid and dashed curves is due
to the different behavior of the respective densities at small
distances $r \le 1$ fm: see fig.\,2, top, where the dashed curve
is $\rho^{RKB}_{sp}(r)$ by eq.(54), while the solid is determined
by
\begin{equation}
\rho^{RKB}(r) = \frac{1}{2\pi^2r}\int\limits_0^\infty F^{RKB}(q)
\sin(qr)qdq.
\end{equation}
Moreover, we see that each of the distributions (for the simple
HOM orbital on the bottom and for the RKB orbital on the top),
after being CM corrected, increases in its central but decreases
in its peripheral region. One may say that we encounter a specific
effect of shrinking the OBDD owing to the translationally
invariant treatment.

In addition, there is a central depression of the density
distribution (cf. the upper and lower dashed lines in fig.2). Such
a change is not unexpected since the RKB WF represents a simple
way to allow for some of the effects of short-range repulsion
between the nucleons in $^{4}He$. These numerical results get an
explicit confirmation if one writes
\begin{equation}
\rho^{RKB}_{sp}(0)=\frac{1}{(\sqrt{\pi}b_H)^3}\frac{1}{1+\beta^2}
\left(1+\sqrt{\frac{3}{2}}\beta\right)^2.
\end{equation}
Evidently, the inequality
\[
\rho^{RKB}_{sp}(0) \le \rho^{RKB}_{sp}(0)\mid_{\beta=0}~
\equiv\rho^{HOM}(0)=\frac{1}{(\sqrt{\pi}b_{H})^3}
\]
takes place for negative $\beta$ values with
$\mid\beta\mid<2\sqrt{6}$.

In parallel, we show in fig. 3 that the corresponding change of
the OBMD has much in common with that for the OBDD, \emph{viz}.,
the distribution $\eta^{RKB}_{EST}(p)$ turns out to be shrunk in
the above sense relative to the distribution $\eta^{RKB}_{sp}(p)$.
Thus, we see a simultaneous shrinking of the density distribution
$\rho(r)$ and the momentum distribution $\eta(p)$. As has been
shown in \cite{DOS76} (see also \cite{Dem88}), such a simultaneous
change of these distributions plays a substantial role in getting
a fair treatment of the data on the elastic and inelastic electron
scattering of $^{4}He$. Let us recall that there the charge FF and
the dynamic FF of $ ^{4}He $ were calculated using one and the
same HOM WF, corrected both with the fixed-CM approximation and
the Peierls-Yoccoz prescription \cite{PY}.

Regarding the properties of these simultaneously corrected
distributions in detail, we would like to emphasize a practical
consequence of their interpretation. This aspect becomes
especially transparent in the case of the simple HOM $(1s)^4$
configuration, where we have
\[
{\rho}_{EST}^{HOM}(r)=[\sqrt{\pi} \bar{r}_0]^{-3}\exp(-
r^2/\bar{r}_0^2)
\]
vs
\[
{\rho}_{sp}^{HOM}(r)=[\sqrt{\pi} r_0]^{-3}\exp(- r^2/r_0^2)
\]
and
\[
{\eta}_{EST}^{HOM}(p)=[\sqrt{\pi} \bar{p}_0]^{-3}\exp(-
p^2/\bar{p}_0^2)
\]
vs
\[
{\eta}_{sp}^{HOM}(p)=[\sqrt{\pi} p_0]^{-3}\exp(- p^2/p_0^2)
\]
Thus, the inclusion of CM corrections gives rise to the two
independent renormalizations, $r_0 \equiv b_H \to \bar{r}_0 =
\sqrt{3/4}r_0 $ and $p_0 \equiv {b_H}^{-1} \to \bar{p}_0 =
\sqrt{3/4}p_0 $, of the oscillatory parameter values, $r_0$ and
$p_0 $ (cf. \cite{DOS76}). Evidently, such changes are not
equivalent to a hasty replacement of $p_0 $ by $\sqrt{4/3}p_0 $ if
one follows the Tassie-Barker recipe with $b_H \to \sqrt{3/4} b_H$
only.

Now, following the conventional way of determining the HOM
parameter $r_0$, as in ref. \cite{DOS76}, we will use the
expansions
\[
F_{ch}(q^2) = 1 - \frac16 q^2 r_{ch}^2 + \cdots ,
\]
\[
f_{p}(q) = 1 - \frac16 q^2 r_p^2 + \cdots
\]
and
\[
F(q) = 1 - \frac16 q^2 r_{rms}^2 + \cdots ,
\]
where we have in HOM
\[
r_{rms}^2 = \frac32 r_0^2,
\]
so that
\[
r_{ch}^2 =  \frac32 r_0^2 + r_p^2,
\]
whence
\begin{equation}
r_0^2 \equiv r_{exp}^2 = \frac{2}{3} \left[r_{ch}^2 - r_p^2
\right].
\end{equation}
Doing so for the CM corrected quantities
we find the similar relation
\begin{equation}
\bar{r}_0^2 = \frac23 \left[r_{ch}^2 - r_p^2 \right] =  r_{exp}^2
\end{equation}
with the identical $q$-dependence $F_{EST}^{HOM}(q) = F^{HOM}(q) =
\exp(-q^2 r_{exp}^2/4 ) $. At the same time the difference between
the respective OBMDs becomes more considerable than after the
substitution $r_0 \to \sqrt{\frac{A}{A-1}} r_{exp}$ in
${\eta}_{sp}^{HOM}(p)=\frac{r_0^3}{ {\pi}^{3/2} } \exp(- p^2
r_0^2) $ that gives
\[
{\eta}_{sp}^{HOM}(p)= \left(\frac{A}{A-1}\right)^{3/2}
\frac{r_{exp}^3}{ {\pi}^{3/2} }\times
\]
\[
\times \exp\left[-\frac{A}{A-1} p^2 r_{exp}^2\right]
\]
vs
\[
{\eta}_{sp}^{HOM}(p)=\frac{r_{exp}^3}{ {\pi}^{3/2} } \exp(- p^2
r_{exp}^2).
\]
Under the simultaneous CM correction of the OBDD and OBMD we have
\[
{\eta}_{EST}^{HOM}(p)=\left(\frac{A}{A-1}\right)^3
\frac{r_{exp}^3}{ {\pi}^{3/2} }\times
\]
\begin{equation}
\times \exp\left[-\left(\frac{A}{A-1}\right)^2 p^2
r_{exp}^2\right]
\end{equation}
vs
\begin{equation}
{\eta}_{sp}^{HOM}(p)=\frac{r_{exp}^3}{ {\pi}^{3/2}} \exp(- p^2
r_{exp}^2),
\end{equation}
that is equivalent to the substitution $r_{0} \to \frac{A}{A-1}
r_{exp} $ in ${\eta}_{sp}^{HOM}(p)=\frac{r_0^3}{ {\pi}^{3/2} }
\exp(- p^2 r_0^2) $.

Note also that the product $\bar{r}_0 \bar{p}_0 = 1-A^{-1} \ne 1
$, unlike the relation
 $r_0 p_0 = 1 $.
In this connection, following \cite{KorShe77} let us remind the
commutation rules for intrinsic coordinate $\;{\vec
r}~^{\prime}={\vec r}-{\vec R}\;$ and conjugate momenta $\;{\vec
p}~^{\prime}={\vec p}-{\vec P}/A\;$
\begin{equation}
[{\vec r}~^{\prime}_{l},{\vec
p}~^{\prime}_{j}]=i{\delta}_{l,j}(1-1/A),\;\;\; (l,j=1,2,3)
\end{equation}
One can show that the corresponding uncertainty principle is
related to the deviation from unity. Thus, the uncertainty
principle does not contradict the simultaneous shrinking of the
density and momentum distributions (see also \cite{She00}, Lect.I,
Suppl. C)

In case of the RKB function we get
\begin{equation}
r_{rms}^2=\frac{3}{2}r_0^2-\frac{\beta\sqrt{6}}{1+\beta^2}
\left(1-\sqrt{\frac{2}{3}}\beta\right)r_0^2.
\end{equation}
It means that the short-range repulsion involved in the WF with a
negative $\beta$ leads to some increasing the rms radius,
\emph{viz.}, $r_{rms}^{RKB}> r_{rms}^{HOM}$. For the values
$\;b_{H}=0.8532$ fm and $\;{\beta}=-0.4738$ the formula $(68)$
yields $r_{rms}^{RKB}=1.429$ fm, so that the corresponding charge
radius is equal to $r_{ch}^{RKB}=1.667$ fm. Here we employ the
charge proton radius $r_p=0.86$ fm (see, for example, Appendix 7
in \cite{ErWei}).

The CM correction gives an opposite effect. Indeed, after some
calculation we find that $r_{rms}^{EST}=1.309$ fm for the same
$\;b_{H}=0.8532$ fm and $\;{\beta}=-0.4738$. From this it follows
that $r_{ch}^{EST}=1.566$ fm.

The variation of $\log {\eta}^{RKB}_{EST}$ and of $\log
{\eta}_{sp}^{RKB}$ with $p$ is depicted in fig.4 for a wider range
of momenta. It is seen from fig.4 that the allowance of the CM
motion improves the description of the available data on the OBMD
of the alpha particle. It is further seen from fig.4 that in the
translationally invariant quantity according to the fixed-CM
prescription the "seagull" \, behavior appearing in the variation
of the corresponding s.p. one becomes somewhat less pronounced.
The dip is diminished and it moves to smaller values of momentum.

Finally, we would like to point out in connection to the
comparison with the s.p. distributions that the CM corrected OBDD
and OBMD become closer to the corresponding microscopic ones by
using their convenient parametrizations from \cite{Morita87} and
\cite{Morita88}, as one can see in figs.3 and 4. According to the
communication \cite{Morita88} one has to introduce the factor
$(2\pi)^{-3}$ to reproduce the momentum distribution
$\eta_{Morita}(p)$ which is one of the significant results
obtained by the Sapporo group. At this point, let us recall that
these authors employed the so-called ATMS-method, where ATMS is
abbreviation "Amalgamation of Two-body correlations into Multiple
Scattering process", to construct the variational WF of the $^4He$
nucleus (see \cite{Akai86} and refs. therein). Along the
variational approach a considerable progress was made when
including more dynamics of the realistic nucleon-nucleon
interaction such as the effect of its tensor component (cf.
\cite{SchiPaWi86}).


\section{Concluding remarks}
We have seen how the approach exposed in  \cite{ShePaMav06} can be
extended to  the translationally invariant evaluation of the
density and momentum distributions in nuclei. The present analysis
shows that the restoration of translational invariance in the
Slater determinant WF of  ${^4}He $  by means of the fixed-CM
correction (the EST prescription) gives rise as a whole to
essential changes in the $r$-, $p$- and $q$-dependences of the
OBDD $\rho (r)$, the OBMD $\eta (p)$ and the charge FF $F_{ch}
(q)$, respectively. We have seen that the correlation between
nucleons induced by the fixation of the center-of-mass of the
nucleus results in the simultaneous shrinking of $\rho(r)$ and
$\eta(p)$. Meanwhile, this effect has been revealed here beyond
the pure HOM extending the available experience.

Also, this study demonstrates the relative importance of the CM
and SRC corrections for the same nucleus, \emph{viz.}, the
shrinking of the density and momentum distributions owing to the
use of translationally invariant g.s. wave functions of $^{4}He$
vs their broadening after the inclusion of short-range repulsion
in these wave functions at small distances $r< 1$ fm. It is true
that the latter has been introduced in our calculations in a
simple manner. Nevertheless, there are all reasons to believe that
the algebraic method employed here might be helpful within more
sophisticated approaches, where the short-range correlations are
taking into account via the Jastrow factor or other correlation
operator (see, {\it e.g.} \cite{Feld98}). At present, the
corresponding applications are in progress both for the $^{4}He$
and $^{16}O$ nuclei.

\vspace{5mm}

{\large Acknowledgements}

\bigskip

One of us, A.S., is very grateful to Michael Grypeos and his
colleagues for their hospitality during his visits to Department
of Theoretical Physics at Aristotle University of Thessaloniki,
where this work was begun. Also, it is a great pleasure for us to
thank A. Antonov and M. Gaidarov for sending the numerical values
of the momentum distribution displayed by the points in fig.4.

\vspace{10mm}

\hspace{5em}{\large Appendix A}

\bigskip

{\bf Some details of calculations beyond HOM}

\bigskip

Here we want to illustrate a convenient method for evaluation of
the expectations in question being aimed at some general
(model-independent) results (cf. \cite{ShePaMav06},
\cite{KorShe77}).

First of all, we have by recurring the Cartesian representation:
\[
\exp{\left[\imath\vec{q}~(\hat{\vec{r}}_1-\hat{\vec{R}})\right]}=
\]
\[
=\exp{\left[\imath\vec{q}~\left(\frac{A-1}{A}\right)~\hat{\vec{r}}_1\right]}
\exp{\left[-\imath\vec{q}~\frac{\hat{\vec{r}}_2}{A}\right]}\dots=
\]
\[
=F_{TB}(q)~F_{HOM}(q)\times
\]
\[
\times\exp{\left[\imath\vec{q}\left(\frac{A-1}{A}\right)\frac{r_0}{\sqrt{2}}~\hat{\vec{a}}_1^{\dag}\right]}\times
\]
\[
\times\exp{\left[\imath\vec{q}\left(\frac{A-1}{A}\right)\frac{r_0}{\sqrt{2}}~\hat{\vec{a}}_1\right]}\times
\]
$$
\times\exp{\left[-\imath\vec{q}\frac{r_0}{\sqrt{2}A}~\hat{\vec{a}}_2^{\dag}\right]}
\exp{\left[-\imath\vec{q}\frac{r_0}{\sqrt{2}A}~\hat{\vec{a}}_2\right]}\dots,
\eqno{(A.1)}
$$
$F_{TB}(q)=\exp(\frac{r_0^2
q^2}{4A})$,~ $F_{HOM}(q)=\exp(-\frac{r_0^2 q^2}{4})$, where the
index $\alpha$ at
$\hat{\vec{a}}_\alpha(\hat{\vec{a}}_\alpha^{\dag})$ is the
individual particle number $(\alpha=1, \cdots, A)$.

Thereat, the Tassie-Barker factor $F_{TB}(q)$ appears
automatically due to a specific structure of the operators
involved. In other words, its appearance is independent of any
nuclear properties (in general, properties of a finite system).
The only mathematical tool that has been used is the
Baker-Hausdorff relation:
$$
\rm{e}^{A+B}=\rm{e}^A~\rm{e}^B~\rm{e}^{-\frac{1}{2}[A,B]},
\eqno{(A.2)}
$$
that is valid with arbitrary operators $\hat{A}$ and $\hat{B}$ for
which the commutator $\left[\hat A,\hat B\right]$ commutes with
each of them. Further, applying eq.(\rm A.2) in combination with
$$
\left(2\pi\right)^3\delta\left(\hat{\vec{R}}\right) =
\int{\exp{\left(\imath\vec{\lambda}\hat{\vec{R}}\right)d\vec{\lambda}}},
\eqno{(A.3)}
$$
one can show that the expectation value $A(q)$ in eq. ($4$) and
the expectation value $N(x)$ in eq. ($16$) are expressed through
one and the same function $F(\vec{x}, \vec{y})$ that depends,
respectively, on the arguments $\vec{x}=\vec{v}$,
$\vec{y}=\vec{s}$ (as in eq.(9)) and $\vec{x}=\vec{v'}$,
$\vec{y}=\vec{s'}$ (as in eq.(18)). In other words, we have
constructed the common generating function for each of them. One
should stress that this result has been obtained independently of
the model WF $\Phi$.

The algebraic technique shown here turns out to be useful for
practical calculations with the Slater determinants like
$\mid\Phi\rangle$ (see \cite {GonShe74}) or the Slater
determinants modified by different correlators (for instance, the
Jastrow factor).

In the simplest case of the independent particle model (IPM)
$(1s)^4$ configuration for $^4He$ with the Slater determinant $
\mid\Phi\rangle=\mid (1s)^4\rangle $ we get omitting the
nonessential factor $[A!]^{-1}$,
$$
A^{IPM}(q)=\exp{\left[-\frac{\bar{r}_0^2
q^{2}}{4}\right]}U^{IPM}(q), \eqno{(A.4)}
$$
$$
U^{IPM}(q)=\int{d\vec{\lambda}\exp{\left[-\frac{r_0^2\lambda^2}{4A}\right]}f(\vec{\lambda},\vec{q}~)},
\eqno{(A.5)}
$$
\[
f(\vec{\lambda},\vec{q})=\langle1s\mid\exp{(-\vec{\alpha}^{*}\hat{\vec{a}}^{\dag})}\exp{(\vec{\alpha}~
\hat{\vec{a}})}\mid1s\rangle\times
\]
$$
\times\langle1s\mid
exp{(-\vec{\beta}^*\hat{\vec{a}}^{\dag})}\exp{(\vec{\beta}\vec{a})}\mid
1s\rangle^3, \eqno{(A.6)}
$$
\[
\vec{\alpha}=\imath\frac{r_0}{\sqrt{2}A}[\vec{\lambda}+(A-1)\vec{q}],
~\vec{\beta}=\imath\frac{r_0}{\sqrt{2}A}[\vec{\lambda}-\vec{q}],
\]
with the renormalized "length" parameter
\[
\bar{r}_0 = \sqrt{\frac{A-1}{A}}~r_0
\]
In our case $r_0 = b_H$. We are keeping the notations with a mass
number $A$ ($=4$) to point out certain trend in the
$A$-dependence.

At this point, let us note that the RKB orbital (or another model
orbital) being composed of the basis states of the spherical
representation can be written then as a superposition of the basis
states  $\mid n_1n_2n_3\rangle$ of the Cartesian representation
(see, \emph{e.g.}, the monograph \cite{NeuSmir} and refs.
therein),
$$
\mid n_1n_2n_3\rangle=\frac{1}{\sqrt{n_1!n_2!n_3!}}
(\hat{\vec{a}}_1^{~\dag})^{n_1}(\hat{\vec{a}}_2^{~\dag})^{n_2}
(\hat{\vec{a}}_3^{~\dag})^{n_3}\mid 0\rangle, \eqno{(A.7)}
$$
where the vector $\mid 0\rangle\equiv\mid 000\rangle$ is the
vacuum state with respect to the destruction operators $\hat{a}_i~
(i=1,2,3) $, {\it e.g}.,
$$
\hat{\vec{a}}\mid 0\rangle=0. \eqno{(A.8)}
$$
It is proved that for the RKB-orbital,
$$
\mid1s\rangle=[1+{\beta}^2]^{-1/2}
[1-({\beta}/{\sqrt{6}})~\hat{\vec{a}}^{~\dag}\hat{\vec{a}}^{~\dag}~]\mid0\rangle.
\eqno{(A.9)}
$$
Substituting ({\rm A.9}) into ({\rm A.6}) (when calculating the
ratio $A^{IPM}(q)/A^{IPM}(0)$, the normalization factor $
[1+\beta^2]^{-1/2}$ can be omitted) we find
$$
\exp{(\vec{\chi}\cdot\vec{a})}\mid
1s\rangle=[1-({\beta}/{\sqrt{6}})(\hat{\vec{a}}^{~\dag}+\vec{\chi})(\hat{\vec{a}}^{~\dag}+\vec{\chi})]\mid0\rangle
\eqno{(A.10)}
$$
for any complex vector $\vec{\chi}$.

Now, after modest effort we obtain
\[
\langle1s\mid
\exp{(-\vec{\chi}~^*\cdot\hat{\vec{a}}^{~\dag})}\exp{(\vec{\chi}\cdot\hat{\vec{a}})}\mid
1s\rangle=
\]
\[
=1+\beta^2-\frac{2}{3}~\beta^2\vec{\chi}~^*\vec{\chi}-
\]
$$
-\frac{\beta}{\sqrt{6}}~[\vec{\chi}~^*\vec{\chi}~^*+\vec{\chi}~\vec{\chi}~]+\frac{\beta^2}{6}~
(\vec{\chi}~^*\vec{\chi}~^*)(\vec{\chi}~\vec{\chi}) \eqno{(A.11)}
$$
From ({\rm A.11}) it follows, for instance,
$$
\langle 1s\mid
\exp{(\imath\vec{\alpha}^*\hat{\vec{a}}^{~\dag})}\exp{(\imath\vec{\alpha}\hat{\vec{a}})}\mid
1s\rangle=M_2\left(\frac{\vec{\alpha}^*\vec{\alpha}}{2}\right),
\eqno{(A.12)}
$$
where the polynomial $M_2(z)$ is given at the end of subsect.
$3.1$.

\end{multicols}
\end{document}